\documentclass[english,aps,prb,preprint,superscriptaddress,showpacs,showkeys]{revtex4}
\usepackage[T1]{fontenc}
\usepackage[latin9]{inputenc}
\usepackage{amstext}
\usepackage{amssymb}
\usepackage{graphicx}

\makeatletter

\providecommand{\tabularnewline}{\\}

\@ifundefined{textcolor}{}
{%
 \definecolor{BLACK}{gray}{0}
 \definecolor{WHITE}{gray}{1}
 \definecolor{RED}{rgb}{1,0,0}
 \definecolor{GREEN}{rgb}{0,1,0}
 \definecolor{BLUE}{rgb}{0,0,1}
 \definecolor{CYAN}{cmyk}{1,0,0,0}
 \definecolor{MAGENTA}{cmyk}{0,1,0,0}
 \definecolor{YELLOW}{cmyk}{0,0,1,0}
 }

\makeatletter
   
\usepackage{amsfonts}

\usepackage{times}

\usepackage{epstopdf}


\makeatother

\makeatother

\usepackage{babel}
\begin{document}

\title{Molecular dynamics modeling of self-diffusion along a triple junction}

\author{T. Frolov}

\email{tfrolov@gmu.edu}

\affiliation{Department of Physics and Astronomy, MSN 3F3, George Mason University,
Fairfax, Virginia 22030, USA}

\author{Y. Mishin}

\email{ymishin@gmu.edu}

\affiliation{Department of Physics and Astronomy, MSN 3F3, George Mason University,
Fairfax, Virginia 22030, USA}
\begin{abstract}
We propose a computational procedure for creating a stable equilibrium
triple junction (TJ) with controlled grain misorientations. We apply
this procedure to construct a TJ between a $\Sigma5(210)$ grain boundary
(GB) and two general high-angle GBs in copper, and calculate the diffusion
coefficients along the TJ and the GBs using molecular dynamics with
an embedded-atom potential. The TJ diffusion is only a factor of two
faster than diffusion in the $\Sigma5$ GB but significantly faster
than diffusion in the general GBs. Both the GBs and the TJ studied
here show a premelting behavior near the bulk melting point, where
their diffusivities converge to the diffusivity of bulk liquid. Although
our results are consistent with the common assumption that TJ diffusion
is generally faster than GB diffusion, the difference between the
two diffusivities does not appear to be large enough to ensure a significant
contribution of TJs to diffusional creep in polycrystals at high temperatures. 
\end{abstract}


\maketitle

\section{Introduction}

The triple junction (TJ) is a line where three grain boundaries (GBs)
meet together. It is commonly believed that TJs greatly influence
properties of nanocrystalline materials and can play a significant
role during plastic deformation and diffusional creep. \citep{Farghalli01,Xiao97,Schuh07}

The TJs are much less studied crystalline defects than GBs or dislocations
are. Experimental investigations of TJs are complicated by many factors,
including their high mobility. \citep{King05} Experimental measurements
indicate that zinc diffusion along TJs in Al is three orders of magnitude
faster than along GBs. \citep{Bokstein01} TJs can also be the sites
of preferred solute segregation. \citep{Yin97} Some theoretical models
introduce the assumption of fast TJ diffusion to explain the anomalous
diffusion and deformation behavior of nanocrystalline materials. \citep{Schuh07,Klinger97,Fedorov02}
However, no reliable experimental information on self-diffusion along
TJs is currently available to verify this assumption.

Due to the recent progress in the development of interatomic potentials
capturing basic properties of materials on the quantitative level,
\citep{Mishin.HMM} atomistic simulations offer a means of predicting
self-diffusion coefficients that would otherwise be very difficult
to measure. For example, the GB self-diffusion coefficients in copper
predicted by simulations were found to be in very good agreement with
experiment. \citep{Suzuki05a} To the best of our knowledge, TJ diffusion
has never been studied by atomistic simulations.

As in experiments, the TJ motion at finite temperatures poses a major
challenge to diffusion simulations. In a few studies published so
far, the TJ motion was prevented by creating high-symmetry configurations
\citep{Srinivasan99} or restricting the simulations to the 0 K temperature.
Applying these restrictions, the atomic structure of the TJs at 0
K was carefully examined and their excess energy was estimated; \citep{Srinivasan99,Shenderova99}
however, no diffusion coefficients were computed. TJ motion by capillary
forces at high temperatures has been studied by two-dimensional Lennard-Jones
simulations. \citep{Upmanyu02}

In this work we apply three-dimensional molecular dynamics (MD) simulations
with an embedded-atom potential to compute self-diffusion coefficients
along a representative TJ and to compare them with self-diffusion
coefficient in the adjoining GBs. Since self-diffusion is the only
type of diffusion studied here, we will be referring to it for brevity
as simply {}``diffusion''. In Section \ref{sec:Method} we show
how a TJ with controlled grain misorientations can be created in the
computer and how its motion can be constrained by boundary conditions.
We also introduce our methodology for diffusion calculations. Our
results are presented in Section \ref{sub:Results}, followed by their
discussion and conclusions in Section \ref{sec:Discussion}.

\section{Methodology\label{sec:Method}}

We chose copper as the model material, with atomic interactions described
by the embedded-atom potential developed in Ref.~\onlinecite{Mishin01}.
The same potential was used in the recent study of GB diffusion \citep{Suzuki05a}.
The bulk melting point of Cu predicted by this potential is $T_{m}=1327$
K (the experimental value is 1356 K).

Our procedure for constructing an equilibrium TJ is illustrated in
Figure \ref{scem}. First, a symmetrical tilt $\Sigma5(210)$ boundary
was created with the tilt axis parallel to the common $\left[100\right]$
direction of both grains. The grain dimensions are $85\times83\times66$
$\textrm{\AA}$, where the third number $\textrm{is the dimension normal to the page}$.
The GB was terminated at a free surface perpendicular to the tilt
axis. Another $116\times66\times66$ $\textrm{\AA}$ grain with  the
orientation shown in Fig.~\ref{scem}   was placed on the surface,
forming a non-equilibrium TJ. The model was equilibrated by a 2 ns
long MD run at 1100 K with a periodic boundary condition parallel
to the TJ and fixed boundary conditions at the bottom and the sides
of the block (the top side is exposed to vacuum). The fixed layers,
shown in Fig.~\ref{scem} in gray, were 10 $\mathring{A}$ thick.
 Further details of the MD methodology will be discussed later.

During the MD anneal, the TJ quickly (within the first 0.1 ns) moved
down, causing migration of the two initially horizontal GBs until
an equilibrium configuration was reached. The planes of the moving
GBs changed in this process, but the grain misorientation remained
the same. The new GB planes were symmetrical with respect to the $\Sigma5$
GB by construction, but they did not align with any particular low-index
crystal planes in the grains. These new boundaries will be referred
to as general GBs (GGBs).

Once equilibrated, the TJ and the GBs did not move except for small
thermal fluctuations. Note that the $\Sigma5$ GB is pinned by the
fixed region at the bottom while the GGBs are pinned by the surface
grooves formed in the corners between the third grain and the two
initial grains.

Finally, a cylindrical region with the axis parallel to the TJ was
cut out of the tricrystal (Fig.~\ref{scem}(c)). The atoms within
a 10 $\mathring{A}$ thick outer shell of the cylinder were made fixed,
whereas all atoms inside the shell were allowed to move during the
subsequent MD simulations. The fixed shell served to pin the GBs and
prevent their migration. Note that the cross-section of the shell
is not perfectly circular: it contains the previously formed surface
grooves and a portion of the bottom fixed layer from the original
block. The boundary condition parallel to the TJ remained periodic.
This cylindrical block contained 81,137 atoms and had the length of
66 $\textrm{\AA}$. It was used as the initial configurations for
the diffusion calculations.

The MD simulations were performed in the $NVT$ (canonical) ensemble
using a Nose-Hoover thermostat. The integration of the equations of
motion employed the Verlet algorithm with the integration time step
of 2 fs. The simulations covered the temperature range from 700 K
(the lowest temperature of reliable diffusion calculations accessible
with our computational resources) to 1315 K (12 K below $T_{m}$).
Prior to an MD run, the block was uniformly expanded/contracted according
to the thermal expansion factor at the intended simulation temperature,
using the thermal expansion factors computed in Ref.~\onlinecite{Mishin01}.
The model was then additionally equilibrated by a 1 ns long MD run
to approach the equilibrium point defect concentrations at the chosen
temperature, followed by a 10 ns long production run for diffusion
calculations. As a test of equilibration, the total energy of the
simulation block was averaged over several time intervals covering
the production stage, and it was checked that these average values
followed a normal distribution. Similar tests were made for the pressure
inside the block. An additional proof of sufficient relaxation comes
from the observations that (i) during the subsequent diffusion runs
(see below), the TJ displacements were random and very small (< 5
$\textrm{\AA}$), and (ii) the mean-squared atomic displacements accurately
followed the Einstein relation. Snapshots containing the atomic coordinates
were saved every 0.02 ns and later post-processed to compute the diffusion
coefficients. 

The diffusion coefficients $D$ were calculated from the linear fits
of $\ \langle X^{2}\rangle$ versus $\Delta t$ plots using the Einstein
relation $D=\langle X^{2}\rangle/2\Delta t$ . Here, $\Delta t$ is
the diffusion time and $\langle X^{2}\rangle$ is the mean-squared
displacement of atoms over the time $\Delta t$ in the $X$ direction
parallel to the TJ. Since the atoms are labeled, $\langle X^{2}\rangle$
was computed by comparing the initial and final coordinates of all
atoms initially contained within a selected probe region and averaging
their squared displacements in $X$ over the time interval between
$t$ and $t+\Delta t$. This obtained value of $\ \langle X^{2}\rangle$
was additionally averaged over all choices of the reference time $t$
by increasing it from zero to $t_{max}-\Delta t$, where $t_{max}$
is the duration of the entire MD run (10 ns). The maximum diffusion
time $\Delta t$ was chosen to be 1 to 2 ns, depending on the temperature.

\begin{figure}[h]

\begin{centering}
\includegraphics[clip,scale=0.6]{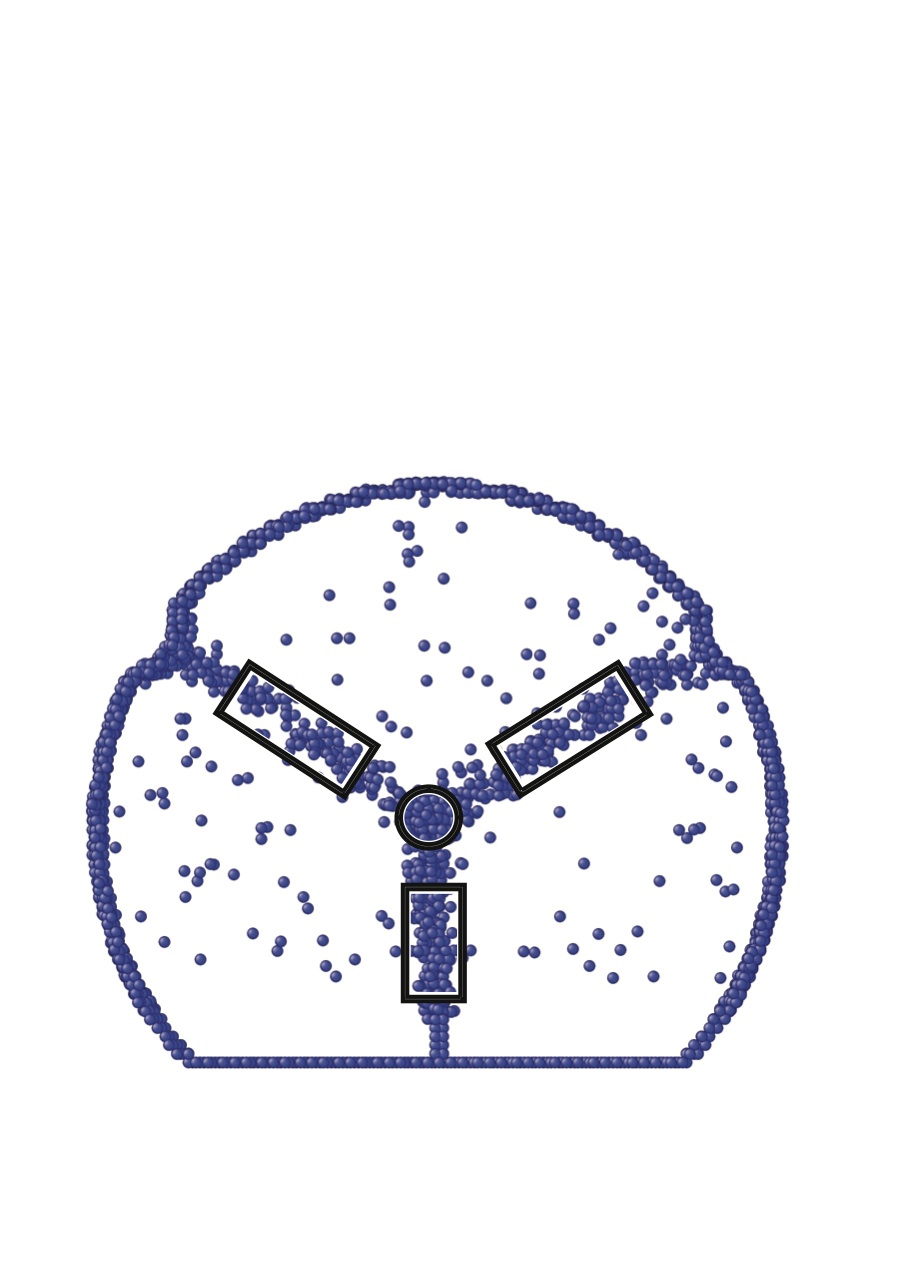} 
\par\end{centering}

\caption{The plot indicates the probe regions chosen for diffusion calculations
for the TJ and the GBs. Only atoms with coordination number below
a certain bound are shown. }

\centering{}\label{scem_reg} 
\end{figure}

The probe regions selected for GB diffusion were parallelepipeds aligned
with the GB planes (Fig.~\ref{scem_reg}), which had the dimensions
$L\times W\times2d$, where $L=66$ $\mathring{A}$ is the cylinder
length. The width $W=35$ $\mathring{A}$ parallel to the GB was chosen
so that to include as large a GB area as possible but avoid the influence
of the TJ and the fixed regions. Since our goal was to evaluate the
diffusivity in the GB core area, the half-thickness of the probe region
was $d=5$ $\mathring{A}$, which is close to the typical diffusion
width of GBs. \citep{Kaur95}

For TJ diffusion, the probe region was a cylinder with the radius
of $d$ placed at the diffusion center of the TJ (Fig.~\ref{scem_reg}).
Special care was taken to keep track of the position of the center.
To achieve this, a testing MD run was made prior to the production
run at each temperature and a group of atoms with the largest magnitude
of displacements in $X$ was identified. Such {}``fastest'' atoms
were invariably found in the TJ region, which was already a qualitative
proof of its enhanced diffusivity. The average coordinates of the
{}``fastest'' atoms were used to select the diffusion center of
the TJ. Furthermore, the position of the center was recomputed for
each snapshot during the production simulations and it was found to
change only by small amounts that did not exceed $d$. This confirmed
the hight stability of our TJ geometry. However, the position of the
center had to be slightly re-adjusted for each temperature since changes
in the contact angles of the GBs could produce TJ displacements.

Diffusion coefficients in liquid copper were computed separately from
mean-squared displacements of atoms at several temperatures using
a large periodic supercell at zero pressure.

\section{Results\label{sub:Results}}

The TJ and GB structures were examined after {}``quenching'' from
700 K to 0 K. Although the structures contained a number of quenched-in
point defects, the were very coherent and had thicknesses on the order
of 2-3 interatomic distances. These structures remained quite coherent
also at finite, but not too high, temperatures. From the dihedral
angle of the GBs, the 0 K energy of the GGB is estimated to be approximately
0.9 of the $\Sigma5$ GB energy. This number is very reasonable, given
that the $\Sigma5(210)$ GB has the highest energy among all boundaries
studied in Refs.~\onlinecite{Suzuki03a,Suzuki03b} and is roughly
10\% above the energies of typical high-angle GBs.

Examination of the MD snapshots reveals that the GBs and the TJ become
increasingly disordered with temperature, as illustrated in Fig.~\ref{tj_struct}.
(The visualization employs the coordination-number mode of the AtomEye
program, \citep{AtomEye} plotting only atoms whose coordination number
is smaller than a certain number.) Near the bulk melting point the
GBs become increasingly wider, especially in the TJ region, and develop
a liquid-like structure. At 1315 K the TJ essentially turns into a
liquid pipe.

\begin{figure}[h]

\begin{centering}
\includegraphics[scale=0.8]{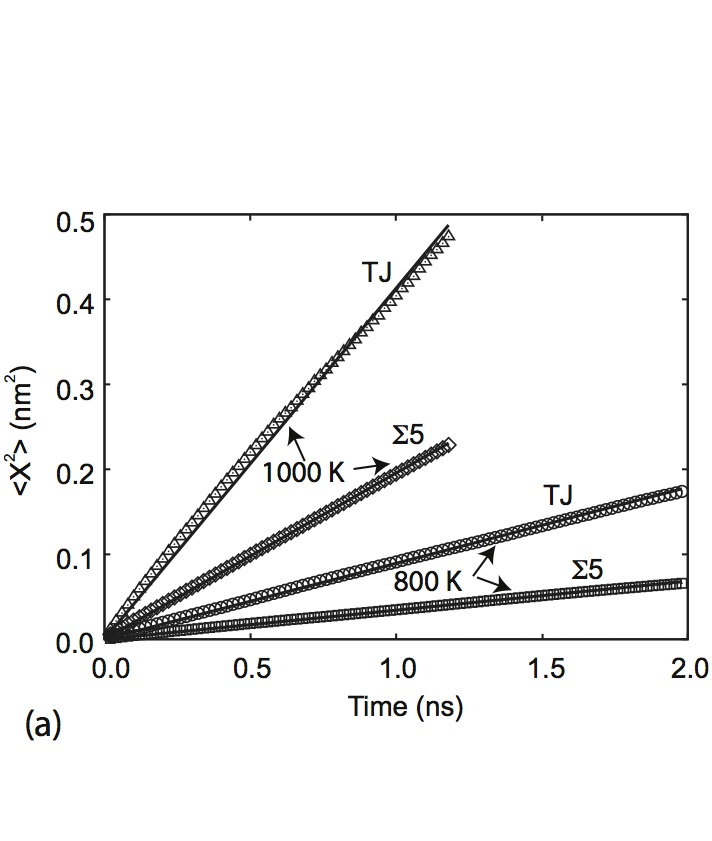} 
\par\end{centering}

\begin{centering}
\includegraphics[scale=0.8]{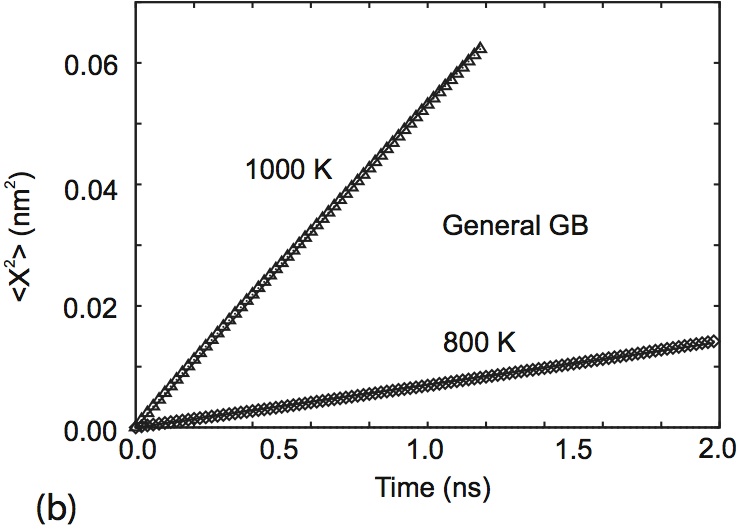}
\par\end{centering}

\caption{Typical plots of the mean-squared displacements calculated for (a)
the $\Sigma5$ GB and the TJ, and (b) the general GB  at 800 and 1000
K. The lines show the linear fits. }

\centering{}\label{msqd} 
\end{figure}

Typical $\langle X^{2}\rangle$ versus time plots are displayed in
Fig.~(\ref{msqd}). Their linearity demonstrates that the Einstein
relation is followed accurately, confirming that we have properly
sampled the diffusion regime. The slopes of the curves indicate that
the TJ diffusion is faster than diffusion in the $\Sigma5$ GB, and
that both diffusivities increase with temperature.

All diffusion coefficients obtained in this work are summarized in
the Arrhenius diagram in Fig.~\ref{Arrh}. Observe the encouraging
agreement with experimental data for self-diffusion in the bulk liquid
\citep{Henderson} and in high-purity polycrystalline Cu, \citep{Surholt97a},%
\footnote{The GB diffusion coefficients were obtained by dividing the measured
products $D\delta$ by the GB diffusion width $\delta=10$ $\mathring{A}$. %
} lending confidence to our simulation methodology. Diffusion along
the TJ is seen to be faster than diffusion in both GBs at all temperatures
tested. Also, the $\Sigma5(210)$ GB shows a higher diffusivity than
the GGB at all temperatures. The latter observation is consistent
with the previous work \citep{Suzuki05a}, in which diffusion in the
$\Sigma5(210)$ GB was found to be faster than in the $\Sigma5(310)$
and $\Sigma17(530)$ GBs. Note the GB and TJ diffusivities converge
at high temperatures and tend the diffusivity of bulk liquid, which
is in agreement with the premelting process. To put these data in
perspective, the self-diffusion coefficient in the lattice computed
with the same EAM potential at the bulk melting point is as low as
$1.1\times10^{-13}$ m$^{2}$/s, which is beyond the scale of Fig.~\ref{Arrh}. 

Despite the significant scatter of the data points, the Arrhenius
plots in Fig.~\ref{Arrh} can be linearized except near the melting
point. A fit according to the Arrhenius equation $D=D_{0}\mbox{exp}\left(-Q/kT\right)$
($k$ is the Boltzmann factor) in the temperature interval 700-1100
K gives the activation energies $Q$ and pre-exponential factors $D_{0}$
summarized in Table \ref{tab:Arrhen}. Although these Arrhenius parameters
slightly depend on the choice of the temperature interval, their ordering
remains the same. The activation energy of TJ diffusion is always
the smallest and close to that of the $\Sigma5$ GB.

\section{Discussion and conclusions\label{sec:Discussion}}

\begin{figure}[h]

\begin{centering}
\includegraphics[scale=0.6]{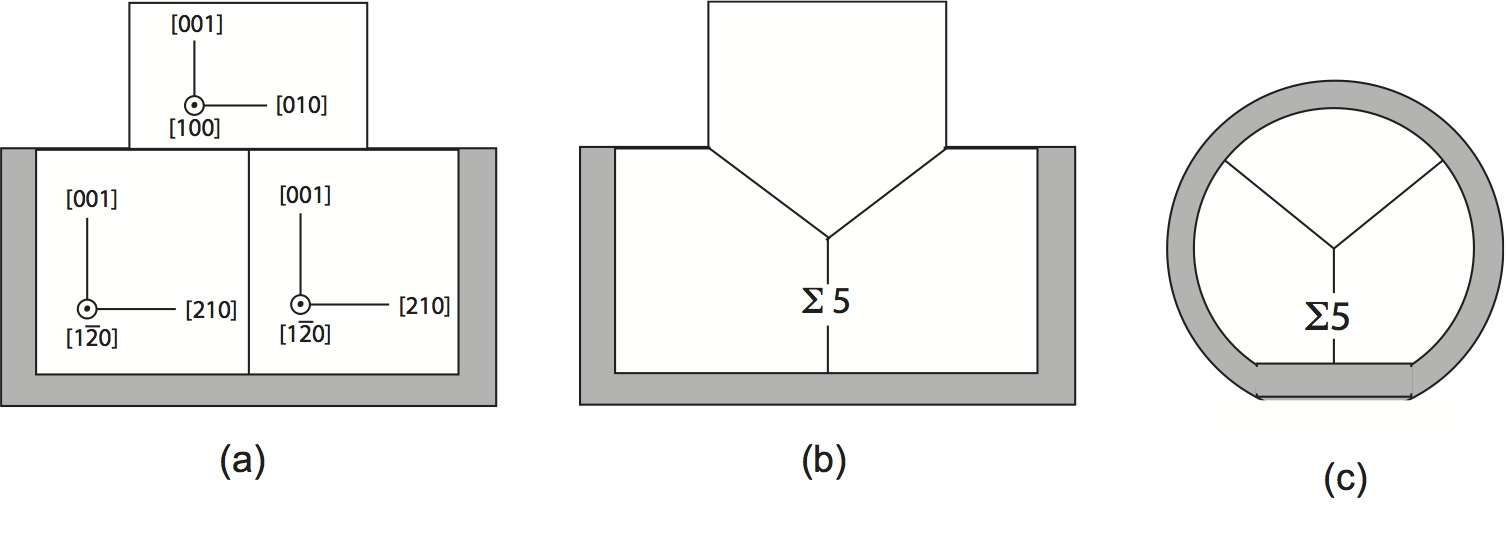} 
\par\end{centering}

\caption{Schematic showing the procedure for the creation of a TJ. (a) Three
grains are joined together by putting a third grain on top of a bicrystal.
(b) An equilibrium TJ forms during an MD run at a high temperature.
(c) A cylindrical region containing the TJ is cut out of the tricrystal
and additionally equilibrated. The shaded areas represent fixed atoms. }

\centering{}\label{scem} 
\end{figure}

We have proposed a computational method to create a single equilibrium
TJ with controlled crystallographic orientations of the grains. The
geometric stability of the TJ is ensured by the fixed boundary conditions
which pin the GBs at three points. An alternative procedure could
be to prepare three separate wedge-shaped grains with angles that
sum to 180$^{\circ}$and join them together like a cake. Besides being
much more tedious, this approach does not offer an easy way to eliminate
the mismatch stresses that may arise due to possible changes in the
translational states of the grains during the relaxation. In addition,
the free volumes of the GBs and of the TJ would have to be accommodated
by elastic stresses. In contrast, in our method the TJ forms \emph{naturally},
as a result of GB migration at a high temperature (Fig.~\ref{scem}(b)).
In this process, the equilibrium atomic density in the GBs and in
the TJ can be reached by atomic diffusion to/from the free surface.
Although in this paper we have implemented a particular set of grain
misorientations, shown in Fig.~\ref{scem}, other  grain misorientations
could have been created as well. A precise control the GB planes is
more difficult, although some control is possible through judicious
choices of the dimensions of the initial grains.

\begin{figure}
\begin{centering}
\includegraphics[scale=0.8]{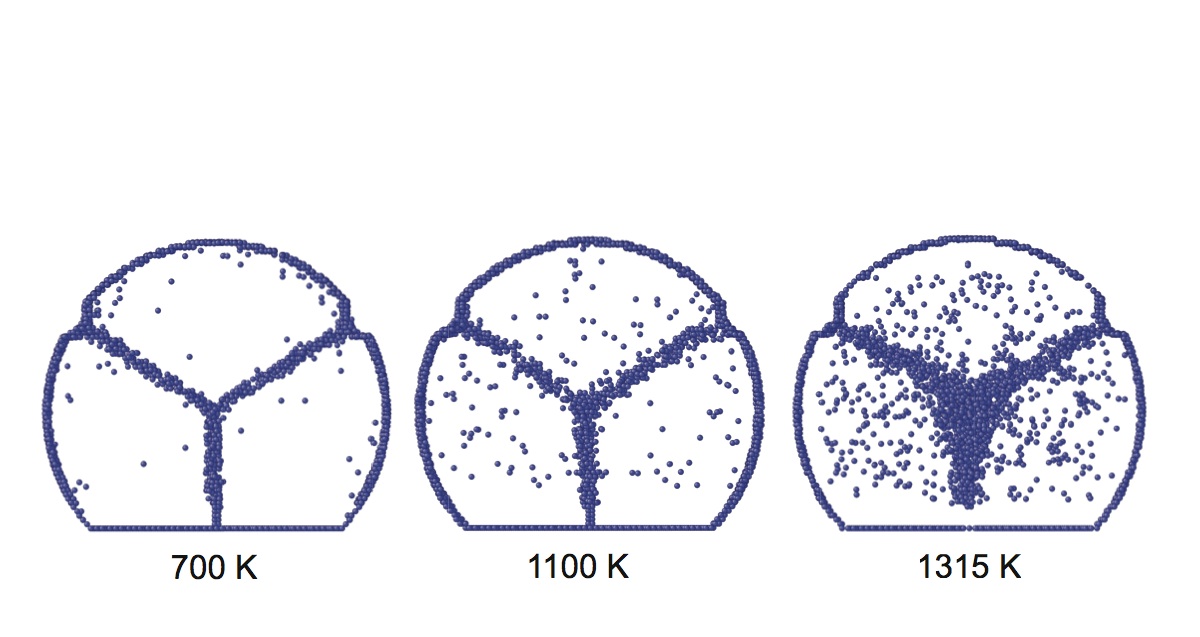} 
\par\end{centering}

\caption{Change of the structure of the GBs and the TJ with temperature. The
visualization method is explained in the text. The bulk melting temperature
with this interatomic potential is 1327 K. }

\centering{}\label{tj_struct} 
\end{figure}

The MD simulations indicate that both the GBs and the TJ become increasingly
disordered at high temperatures and exhibit a premelting behavior
near the bulk melting point (Fig.~\ref{tj_struct}). As the temperature
approaches $T_{m}$, the premelting first starts at the TJ, then extends
to the $\Sigma5$ GB, and finally occurs at the GGBs. This sequence
of premelting processes suggests that the TJ has a higher excess free
energy than the GBs studied here. The question of whether TJs carry
an excess free energy relative to the adjoining GBs ({}``excess of
excesses'') has long been debated in the literature. \citep{Srinivasan99}

Diffusion simulations have been conducted over a wide temperature
range and the diffusion coefficients in the TJ and the GBs have been
extracted from the mean-squared atomic displacements. It should be
mentioned that due to the fixed boundary conditions and conserved
number of atoms in the model, the point-defect equilibrium at the
diffusion temperature could only be achieved by defect generation
and/or absorption at the TJ and the GBs. At high temperatures, the
TJ and the boundaries were premelted or at least highly disordered,
making them effective sinks and sources of point defects. Even at
the lowest temperature studied here (700 K), the TJ region was still
fairly disordered and was presumably able to generate and absorb point
defects during the pre-diffusion anneal. This was evidenced by the
linearity of the mean-square displacements versus time plots at this
temperature, as well as by the independence of the diffusion results
on the thermal history. For example, an additional low-temperature
anneal of the TJ prior to carving out the cylinder (i.e., when the
GBs were still connected to the surface, Fig.~\ref{scem}(b)) did
not change the diffusion coefficients within the statistical error.
It is important to recognize, however, low-temperature MD simulations
in closed systems containing of well-ordered GBs or other extended
defects can be associated with non-equilibrium point-defect concentrations
and can give inaccurate diffusion coefficients. 

\begin{figure}[h]
\begin{centering}
\includegraphics[scale=0.75]{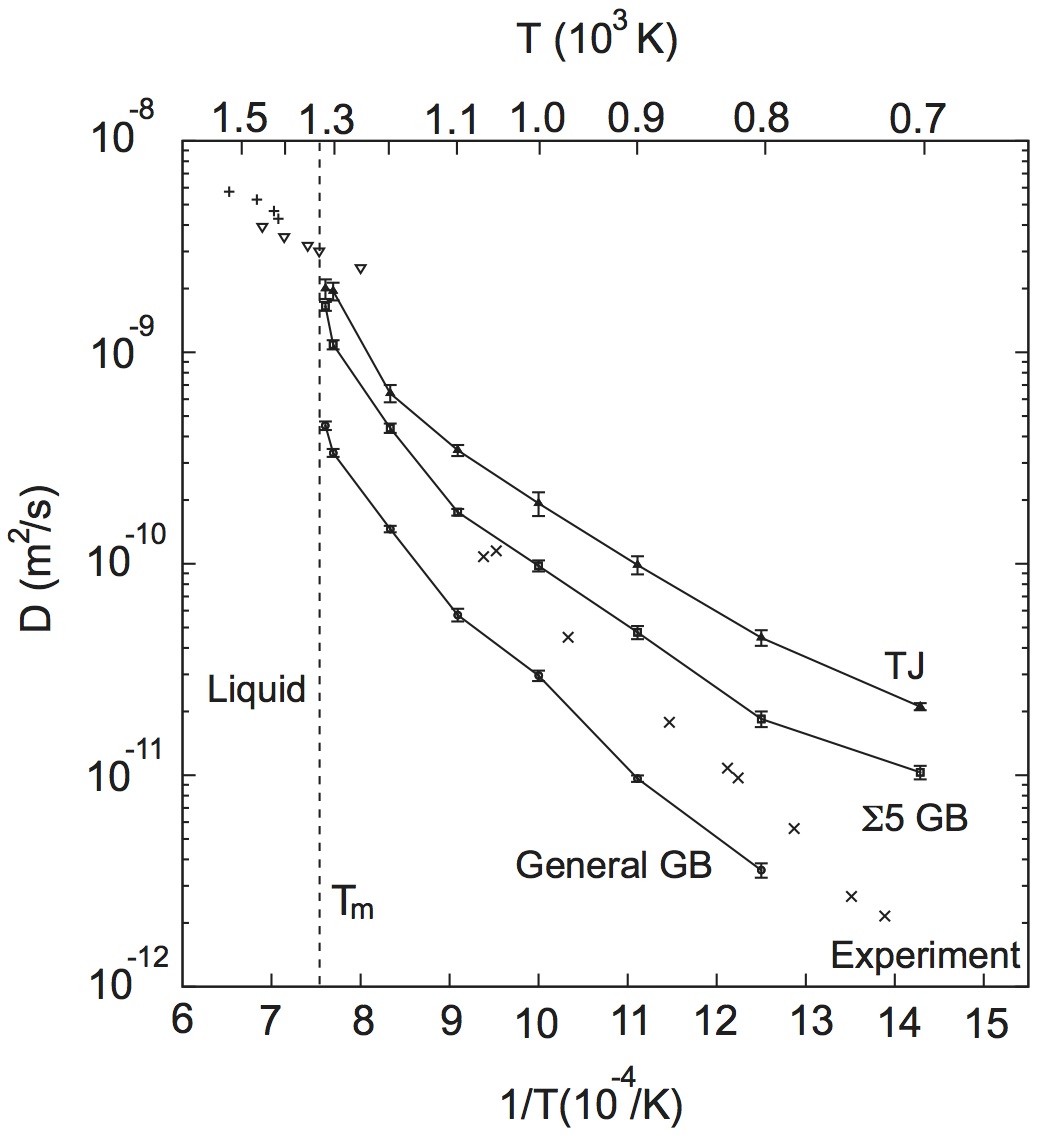} 
\par\end{centering}

\caption{Arrhenius diagram of diffusion coefficients $D$ calculated for the
TJ ($\triangle$), the $\Sigma5$ GB ($\square$), the GGB ($\bigcirc$)
and liquid copper ($\nabla$). The vertical dashed line indicates
the melting point of Cu with this embedded-atom potential (1327 K).
Experimental data for self-diffusion in liquid copper\citep{Henderson}
($+$) and in high-purity copper polycrystals\citep{Surholt97a} ($\times$)
are shown for comparison.}

\centering{}\label{Arrh} 
\end{figure}

Our calculations confirm that diffusion along the TJ is faster than
diffusion along the GBs at all temperatures (Fig.~\ref{Arrh}). Our
results also suggest that, for diffusion along {}``typical'' GBs,
such as the GGB tested here, can be orders of magnitude slower than
TJ diffusion at low temperatures. This conclusion is consistent with
recent experimental measurements. \citep{Bokstein01} However, diffusion
in the highly energetic $\Sigma5(210)$ GB is only a factor of two
slower than in the TJ. Although this small gap tends to slightly increase
at low temperatures, Fig.~\ref{Arrh} and the Arrhenius parameters
(Table \ref{tab:Arrhen}) clearly indicate that high-diffusivity GBs
such as $\Sigma5$ can be more efficient for atomic transport in polycrystals
than TJs. The factor of two higher diffusivity of TJs is hardly sufficient
for making a significant contribution considering their much smaller
cross-sectional area in comparison with GBs.

As the temperature approaches the melting point, the TJ and GB diffusivities
converge to the bulk liquid diffusivity, reflecting the premelting
process. A very similar behavior was recently observed for GB diffusion.
\citep{Suzuki05a} The rapid increase of the diffusivity near $T_{m}$
produces a strong upward curvature of the Arrhenius plots and the
concept of activation energy loses its physical significance. From
the convergence of the diffusivities at high temperatures, it appears
unlikely that TJs can make a sensible contribution to diffusional
creep at high temperatures, given their relatively small cross-sectional
area even in nanocrystalline materials. With most GBs in nanocrystals
being of general type, the contribution to the creep rates often attributed
to TJs might be overestimated.

\bigskip{}

\textbf{Acknowledgments }

We are grateful to W.~J.~Boettinger for numerous fruitful discussions.
We have benefited greatly from interactions with many colleagues during
coordination meetings sponsored by the DOE-BES Computational Materials
Science Network (CMSN) program. This work was supported by the US
Department of Energy, Office of Basic Energy Sciences. We used the
resources of the National Energy Research Scientific Computing Center
supported by the Office of Science of the US Department of Energy
under Contract No.~DE-AC02-05CH11231.

\bigskip{}

%\bibliographystyle{/Users/ymishin/YURI/Bibliography/aip} \bibliographystyle{/Users/ymishin/YURI/Bibliography/aip}
%\bibliography{/Users/timfrol/Articles/Biblography/literat}

\bigskip{}

\bigskip{}
 \newpage{} \clearpage{}

\begin{table}
\caption{Arrhenius parameters $Q$ (activation energy) and $D_{0}$ (pre-exponential
factor) of TJ and GB diffusion in Cu obtained in this work. The Arrhenius
parameters of lattice self-diffusion computed with the same interatomic
potential are given for comparison. \label{tab:Arrhen}}

\bigskip{}

\begin{centering}
\begin{tabular}{|l|c|c|}
\hline 
Defect  & $Q$ (eV)  & $D_{0}$ (m$^{2}$/s)\tabularnewline
\hline 
\hline 
Triple junction  & $0.47\pm0.02$  & $(4.3\pm1.3)\times10^{-8}$\tabularnewline
\hline 
$\Sigma5$ GB  & $0.48\pm0.04$  & $(2.5\pm1.4)\times10^{-8}$\tabularnewline
\hline 
General GB  & $0.71\pm0.02$  & $(1.0\pm0.2)\times10^{-7}$\tabularnewline
\hline 
Perfect lattice & 1.961$^{a}$ & $3.1\times10^{-6}$$^{b}$\tabularnewline
\hline 
\end{tabular}\bigskip{}

\par\end{centering}

\noindent \begin{raggedright}
$^{a}$Sum of the vacancy formation and migration energies (Table
2 in Ref. \onlinecite{Suzuki03a})
\par\end{raggedright}

\noindent \raggedright{}$^{b}$Computed from the lattice parameter
and the vacancy formation entropy and jump-attempt frequency (Table
2 in Ref. \onlinecite{Suzuki03a})\vspace{1.5cm}
\end{table}

%\newpage \clearpage%

%EndExpansion

\end{document}